# Efficient intrinsic spin-to-charge current conversion in an all-epitaxial single-crystal perovskite-oxide heterostructure of La$_{0.67}$Sr$_{0.33}$MnO$_3$/LaAlO$_3$/SrTiO$_3$

Shinobu Ohya,[1,2,3*] Daisei Araki,[1] Le Duc Anh,[1,2] Shingo Kaneta,[1] Munetoshi Seki,[1,3] Hitoshi Tabata,[1,3] and Masaaki Tanaka[1,3]

[1]*Department of Electrical Engineering and Information Systems, The University of Tokyo, 7-3-1 Hongo, Bunkyo-ku, Tokyo 113-8656, Japan*
[2]*Institute of Engineering Innovation, Graduate School of Engineering, The University of Tokyo, 7-3-1 Hongo, Bunkyo-ku, Tokyo 113-8656, Japan*
[3]*Center for Spintronics Research Network (CSRN), Graduate School of Engineering, The University of Tokyo, 7-3-1 Hongo, Bunkyo-ku, Tokyo 113-8656, Japan*

Abstract

We demonstrate efficient intrinsic spin-to-charge current conversion in a two-dimensional electron gas using an all-epitaxial single-crystal heterostructure of LaSrMnO$_3$/ LaAlO$_3$ (LAO)/ SrTiO$_3$ (STO), which can suppress spin scattering and give us an ideal environment to investigate *intrinsic* spin-charge conversion. With decreasing temperature to 20 K, the spin-to-charge conversion efficiency is drastically enhanced to +3.9 nm, which is the largest positive value ever reported for LAO/STO. Our band-structure calculation well reproduces this behavior and predicts further enhancement by controlling the density and relaxation time of the carriers.

*Email: ohya@cryst.t.u-tokyo.ac.jp

Recent observations of conversion phenomena between spin and charge currents promise substantial reduction of power consumption in next-generation high-speed spintronics devices such as spin-orbit-torque magnetoresistive random-access memories [1]. While this conversion is well known to occur in heavy metals [2–4], recent studies have shown that it occurs also at various interfaces such as Ag/Bi, Fe/Ge, and Ag/α-Sn [5−9]. At these interfaces, the Rashba spin-orbit interaction and resulting spin splitting of the Fermi surface appear due to the broken space-inversion symmetry, causing spin-to-charge current conversion that is known as the inverse Edelstein effect (IEE) [10,11].

Very recently, a giant IEE was observed in a two-dimensional electron gas (2DEG) formed at the interface between insulating perovskite oxides $LaAlO_3$ (LAO) and $SrTiO_3$ (STO) [12]. In this heterostructure, carriers are provided from oxygen vacancies and form a high-mobility 2DEG at this interface [13]. The strong Rashba spin-orbit interaction at LAO/STO, which can be modulated by a gate voltage, makes this system very attractive for controllable efficient spin-charge conversion [14–16]; however, previously reported results for spin-charge conversion at LAO/ STO are divergent, and unified understanding of its intrinsic mechanism is still lacking [12,17–19]. For example, in Ref. [12], a large conversion efficiency, so called the inverse Edelstein length $\lambda_{IEE}$, up to −6.4 nm was observed at 7 K. This value is much larger than that reported for Ag/Bi ($\lambda_{IEE} = 0.3$ nm). Meanwhile, in Refs. [17,18], the conversion signal strongly decreases to zero with decreasing temperature. The reason for these completely different temperature dependences is not clear at present, but this is likely attributed to inelastic transport of the spin current, which is predicted to reduce the conversion signal especially at low temperature. This inelastic spin transport is thought to be related to the crystal quality of samples. In metal systems, the interface quality is known to have a large influence on the conversion efficiency [20,21]. In all the previous studies on spin-charge conversion at LAO/ STO [12,17–19], however, amorphous or poly-crystalline ferromagnetic films deposited by sputtering were used for the ferromagnetic layer, which may cause strong spin scattering especially at the interface between the ferromagnetic layer and LAO. In this Letter, to exploit the intrinsic IEE in the LAO/STO system, we focus on an all-epitaxial single-crystal heterostructure of $La_{0.67}Sr_{0.33}MnO_3$ (LSMO)/ LAO/ STO. LSMO is a strongly-correlated half-metallic ferromagnetic-perovskite oxide that can be epitaxially grown on STO due to the small lattice mismatch of ~0.8%. LSMO is thus an ideal candidate to explore the efficient spin injection and intrinsic spin-charge conversion at the LAO/STO interface.

For the experiments, we have prepared a sample composed of $La_{0.67}Sr_{0.33}MnO_3$ [30

unit cell (u.c.) = 12 nm)]/ La$_{(1-\delta)}$Al$_{(1+\delta)}$O$_3$ (LAO, 2 u.c. = 0.8 nm) grown on a TiO$_2$-terminated SrTiO$_3$ (001) substrate via molecular beam epitaxy (MBE) [Fig. 1(a)]. We used a shuttered growth technique with fluxes of La, Sr, Mn, and Al supplied by Knudsen cells. The LAO and LSMO layers were grown at 730°C with a background pressure of $7 \times 10^{-4}$ Pa due to a mixture of oxygen (80%) and ozone (20%). As shown later, the thickness of 2 u.c. of LAO is large enough to form a 2DEG at the LAO/STO interface because of the presence of the LSMO layer as with previous reports on LAO/STO with a metallic capping layer [12,22,23]. As shown in Fig. 1(b), in which we assume that a simple parabolic band structure is split due to the Rashba effect, the spin current that is injected into the LAO/STO interface moves the outer and inner Fermi surfaces in the opposite directions under the ferromagnetic resonance (FMR) condition, generating a charge current in the [$\bar{1}$10] ($x$) direction. This effect induces the electromotive force (EMF) between the electrodes at the edges of the sample in the [$\bar{1}$10] direction [Fig. 1(a)]. We note that the EMF includes a signal originating from LSMO that is induced by the microwave electric field, such as the galvanomagnetic effects (*e.g.* anomalous Hall effect and planer Hall effect), which should be separated from the IEE signal. For this purpose, we have grown two LSMO/ LAO/ STO samples with a 2DEG (sample A) and without a 2DEG (sample B); as shown in a previous study on MBE-grown LAO/STO films [24], a 2DEG is formed only when the composition ratio $c = (1-\delta)/(1+\delta)$ of La to Al in La$_{(1-\delta)}$Al$_{(1+\delta)}$O$_3$ is below $0.97 \pm 0.03$. In samples A and B, $c$ is set at 82.5 % and 101%, respectively. Because the IEE is observed only in sample A, we can extract the pure IEE signal by comparing the results between samples A and B. The scanning transmission electron microscope (STEM) image of sample A (LSMO/LAO/STO) shown in Fig. 1(c) confirms that all the layers are single-crystalline and coherently grown on the STO substrate. The sample surface is atomically flat with atomic steps [Fig. 1(d)]. The LSMO layer has a Curie temperature above room temperature [Fig. 2(a)].

To confirm that a 2DEG is formed only when $c$=82.5%, we measured the transport properties of reference samples of La$_{(1-\delta)}$Al$_{(1+\delta)}$O$_3$ (8 u.c.= 3.2 nm)/STO with $c$=82.5% (named sample ref-A) and $c$=101% (named sample ref-B), which were grown with the same growth conditions as those for samples A and B, respectively. Actually, as shown in Fig. 2(b), sample ref-A ($c$=82.5%) shows metallic behavior while sample ref-B ($c$=101%) shows insulating behavior, confirming that a 2DEG exist only when $c$=82.5%. By the Hall measurements, the sheet career density $n_s$ of the 2DEG was estimated to be $2.1 \times 10^{14}$ cm$^{-2}$ and the mobility $\mu$ was $3.7 \times 10^3$ cm$^2$V$^{-1}$s$^{-1}$ at 20 K [Fig. 2(c)]. As shown in Fig. 2(d), sample A shows similar metallic behavior, while sample B shows nearly the same temperature dependence of the sheet resistance as that of a single LSMO layer grown on STO (dotted curve reproduced from Ref. [25]). These results confirm the presence of the 2DEG only in sample A.

We have carried out spin pumping measurements using a $TE_{011}$ cavity of an electron-spin-resonance system with a microwave frequency of 9.1 GHz. We cut the samples into a small piece with a size of $2 \times 1$ mm and put it at the center of the cavity. For the measurements, a static magnetic field $\mu_0 H$ was applied along the [110] ($y$) direction in the film plane, which corresponds to the easy magnetization axis of LSMO. Meanwhile the microwave magnetic field $h_{rf}$ was applied along the [$\bar{1}10$] direction. The used microwave power was 30 mW.

As shown in Figs. 3(a) and 3(b), the EMF peak appears at the FMR magnetic field at all the measurement temperatures, indicating that the measured EMF is induced by the FMR, like in general spin pumping experiments. We note that we can eliminate the influence of the thermal effects as discussed in Sec. 1 of the Supplemental Material (SM) [26]. To derive the IEE signal from the EMF, we extracted the symmetric component $V_s$, which includes the IEE signal, from the EMF−$H$ curves (Sec. 2 in the SM [26]). Then, we estimated the sheet current density $j_c^A = V_s/(wR)$, where $R$ is the resistance [see Fig. 2(d)] and $w$ is the sample width (1 mm). In Fig. 3(c), one can see a drastic increase in $j_c^A$ with decreasing temperature.

To separate the IEE signal from the one originating from LSMO such as the galvanomagnetic effects, we derived the IEE-induced sheet current density $j_c^{2D}$ by subtracting the sheet current density $j_c^B$, which was measured for sample B (see Sec. 3 of SM [26]), from $j_c^A$. As shown in Fig. 4(a), $j_c^A$ is much larger than $j_c^B$ especially at low temperatures, indicating that $j_c^A$ is mainly attributed to the IEE signal.

Using obtained $j_c^{2D}$ ($= j_c^A - j_c^B$) and the estimated value of the spin current density $j_s$, which is derived from the spectral linewidth of the FMR signals with a standard method described in Sec. 4 of SM [26], we estimated $\lambda_{IEE}$ ($= j_c^{2D}/j_s$) at each temperature [red open circles in Fig. 4(b)]. In Fig. 4(b), $\lambda_{IEE}$ drastically increases with decreasing temperature and amounts up to +3.9 nm at 20 K. This value is the highest positive value ever reported for the IEE at LAO/STO. This temperature dependence is similar to that in Ref. [19] but is completely opposite to that reported in Refs. [17,18].

As discussed below, this characteristic increase in $\lambda_{IEE}$ with decreasing temperature mainly originates from the intrinsic feature of the IEE in the LAO/STO system. Following the approach in Ref. [27], we calculated the band structure of the LAO/STO interface using the effective-mass Hamiltonian with atomic spin-orbit coupling and interorbital nearest-neighbor hopping based on the six $3d$–$t_{2g}$ orbitals of up and down spin components of the $d_{xy}$, $d_{yz}$, and $d_{zx}$ orbitals of Ti (see Sec. 5 in SM [26]). The calculated band structure is shown in Fig. 4(c). Comparing the $n_s$ values obtained for sample ref-A [see Fig. 2(c)] and the theoretical carrier density (Fig. S5 in SM [26]), we estimated the $E_F$ positions in our samples, which are shown as the red dotted lines in Fig. 4(c). From the Boltzmann equation, for the $n$-th Fermi surface SF$_n$, the two-dimensional (2D) current density $j_c^{\mathrm{FS}n}$ and the non-equilibrium spin density $\delta s^{\mathrm{FS}n}$ are expressed by

$$j_c^{\mathrm{FS}n} = \frac{e^2}{4\pi^2\hbar}\int^{\mathrm{FS}n} F_x(\mathbf{k})\,\mathrm{d}S_F, \quad \delta s^{\mathrm{FS}n} = \frac{e}{4\pi^2\hbar}\int^{\mathrm{FS}n} S_y(\mathbf{k})\,\mathrm{d}S_F, \qquad (1)$$

where $e$ is the free electron charge, $\hbar$ is the Dirac constant, $\mathrm{d}S_F$ is the infinitesimal area (= length in two dimensions) of the Fermi surface [28]. Here, $F_x(\mathbf{k})$ and $S_y(\mathbf{k})$ are defined as

$$F_x(\mathbf{k}) = F\,\mathrm{sgn}\left(S_y(\mathbf{k})\right)\tau(\mathbf{k})v_x(\mathbf{k})\frac{v_x(\mathbf{k})}{|\mathbf{v}(\mathbf{k})|}, \quad S_y(\mathbf{k}) = F\tau(\mathbf{k})\sigma_y(\mathbf{k})\frac{v_x(\mathbf{k})}{|\mathbf{v}(\mathbf{k})|}, \qquad (2)$$

where $F$ is the absolute value of the effective electric field that is applied to each electron state, and $v_x(\mathbf{k})$ is the $x$ direction component of the group velocity $\mathbf{v}(\mathbf{k})$. We assumed that the relaxation time $\tau(\mathbf{k})$ is proportional to $|\mathbf{k}|$. Then, $j_c^{\mathrm{2D}}$, the total non-equilibrium spin density $\delta s$, and $\lambda_{\mathrm{IEE}}$ are expressed by

$$j_c^{\mathrm{2D}} = \sum_n j_c^{\mathrm{FS}n}, \qquad \delta s = \sum_n \left|\delta s^{\mathrm{FS}n}\right|, \qquad \lambda_{\mathrm{IEE}} = \frac{j_c^{\mathrm{2D}}}{j_s} = \frac{\tau}{e}\frac{j_c^{\mathrm{2D}}}{\delta s}. \qquad (3)$$

The most important indication in the above equations is that the charge current is mainly carried by electrons with large $v_x(\mathbf{k})$. As shown in the calculated $v_x(\mathbf{k})$ mapping at the Fermi surface when $E_F = 210$ meV [Fig. 4(d)], which corresponds to the case of the measurement temperature of 20 K in our study, we see that electrons in the vicinity of $k_y = 0$ mainly contribute to the charge current. In fact, especially the $d_{xy}$ states that are located at $k_x = \sim \pm 0.186\pi/a$ have large $v_x(\mathbf{k})$ and $F_x(\mathbf{k})$ as shown in Figs. 4(d) and 4(e), where $a$ is a lattice constant of STO. Similarly, when $E_F \geq 210$ meV, the $d_{xy}$ states near $k_y = 0$ have a dominant contribution to the charge current, leading to a nearly energy-independent value of $j_c^{\mathrm{2D}}/\delta s$ (= $e\lambda_{\mathrm{IEE}}/\tau$) when $E_F \geq 210$ meV (Sec. 6 in SM [26]). Thus, $\lambda_{\mathrm{IEE}}$ is almost determined by $\tau$ in this energy region.

In a way similar to the derivation of Eqs. (1)–(3), we can obtain the relaxation time $\tau$

from the experimental sheet resistance of sample A shown in Fig. 2(d) (Sec. 7 in SM [26]). Using the theoretical value of $j_c^{2D}/\delta s$ (see Fig. S6 in SM [26]) and $\tau$, we can predict $\lambda_{\text{IEE}}$ that is expected in our system at each temperature [square points in Fig. 4(b)]. We see that predicted $\lambda_{\text{IEE}}$ increases with decreasing temperature as with the experimental $\lambda_{\text{IEE}}$, which confirms that our result originates from the intrinsic IEE. The reason for the larger values of predicted $\lambda_{\text{IEE}}$ than the experimental values especially at low temperatures is probably due to our overestimation of $j_s$ (see Sec. 4 in SM [26]) or small influence of spin scattering.

Our study shows that the small thickness of LAO only of 2 u.c. and the high-quality single crystallinity are likely keys to suppressing the extrinsic effect and to obtaining intrinsic IEE. Furthermore, our band calculation suggests that $\lambda_{\text{IEE}}$ will be dramatically enhanced if we can tune the $E_F$ position at around the Lifshitz point (see Sec. 6 in SM [26]). At the same time, we see that increasing $\tau$ is important to enhance the IEE, indicating that single crystalline 2D system with a high mobility is very promising for efficient conversion between spin and charge currents.

This work is partly supported by Grants-in-Aid for Scientific Research by MEXT (Grant No. 18H03860), CREST of JST (Grant No. JPMJCR1777), Advanced Characterization Nanotechnology Platform of the University of Tokyo by MEXT, and the Spintronics Research Network of Japan (Spin-RNJ). We thank Mr. Takeshima for technical help for the MBE growth and ESR measurements.

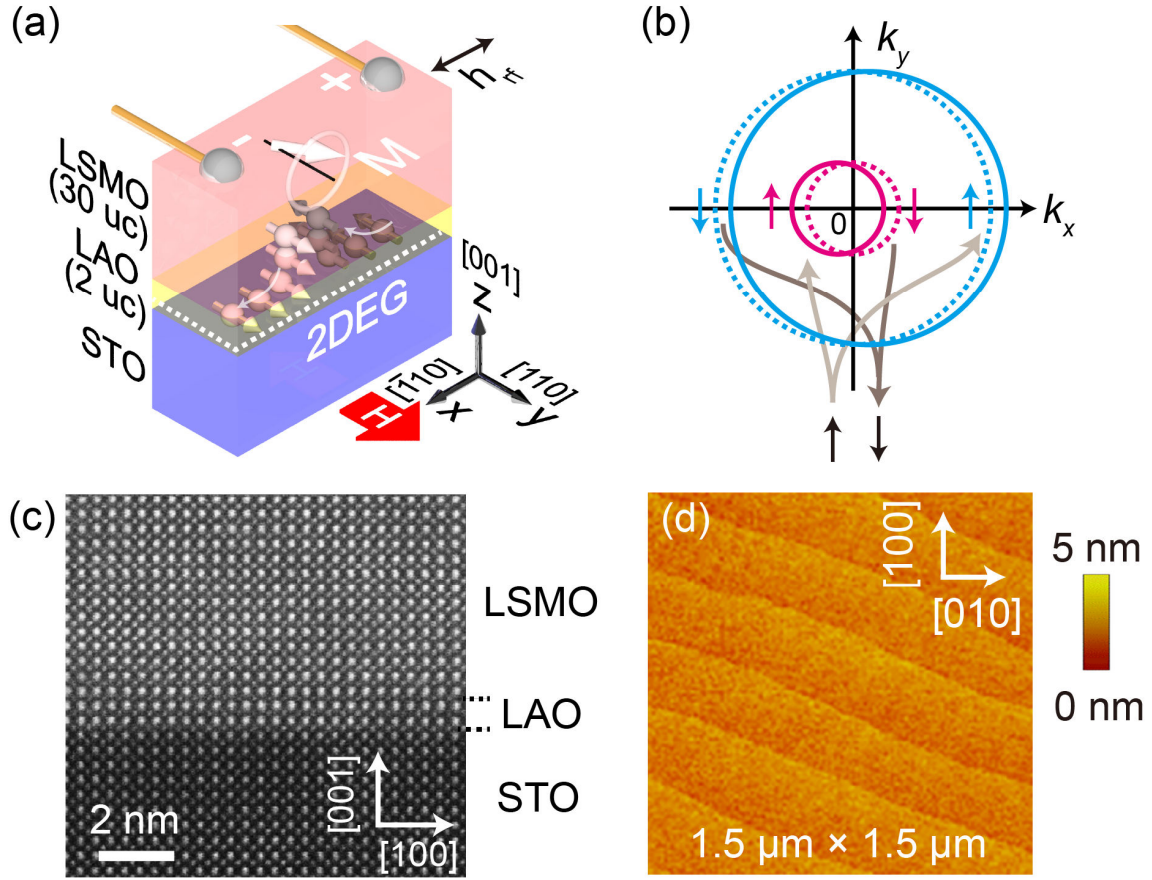

FIG. 1. (a) Schematic illustration of the (001)-oriented full-epitaxial multilayer structure of LSMO/LAO grown on an STO (001) substrate. The sample size is 2×1 mm. In the electron-spin-resonance system, a radio-frequency magnetic field $h_{rf}$ was applied along the $[\bar{1}10]$ ($x$) direction of the sample. The static magnetic field $\mu_0 H$ was applied along the [110] ($y$) axis. Here, $M$ represents the magnetization of LSMO. (b) Principle of spin-to-charge conversion via the inverse Edelstein effect; the spin current injected into the LAO/STO interface moves the outer and inner Fermi circles, generating a charge current in the $x$ direction. Here, the dotted and solid lines are the original Fermi circles and the ones after a spin current is injected, respectively. (c) Scanning-transmission-electron microscope-lattice image of the LSMO (30 u.c.)/ LAO (2 u.c.)/ STO heterostructure (sample A) projected along the [010] axis. (d) Atomic-force-microscope image of the surface of sample A, in which atomic steps are observed.

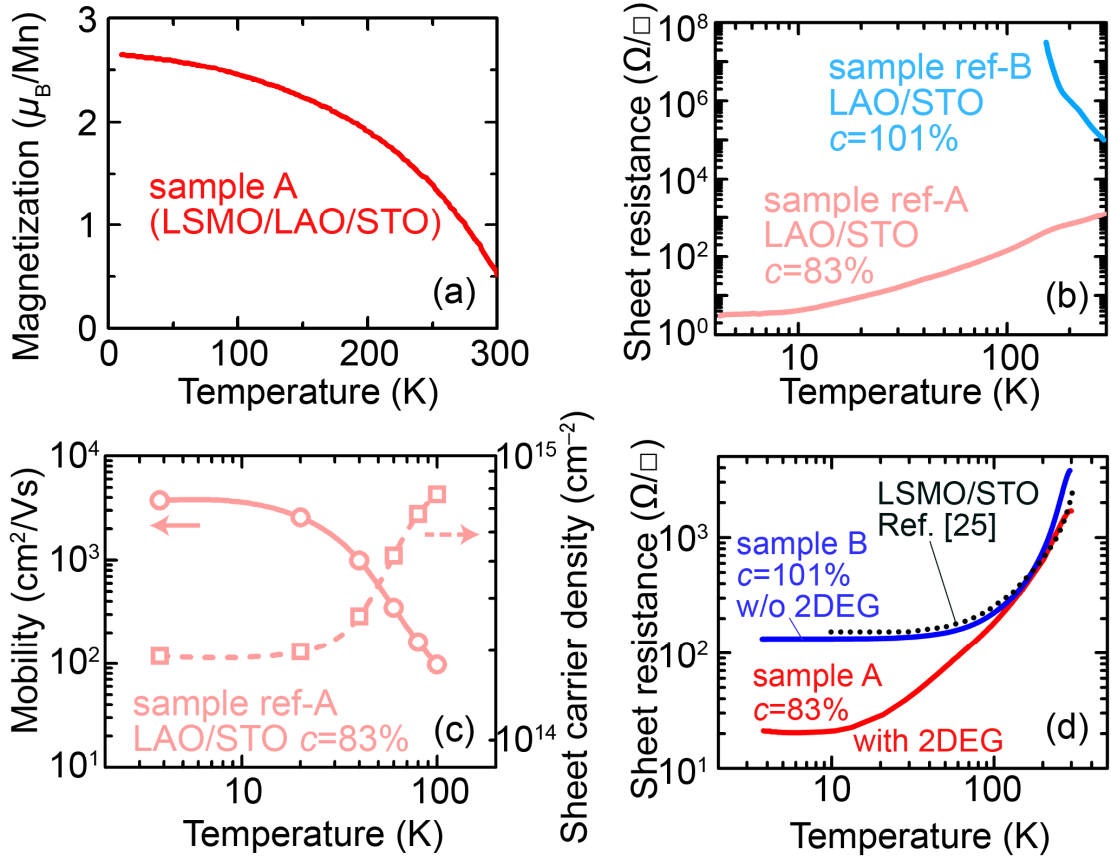

FIG. 2. (a) Temperature dependence of the magnetization of LSMO (30 u.c.)/ LAO (2 u.c.)/ STO (sample A). (b) Temperature dependence of the sheet resistance of the reference samples of LAO (8 u.c.)/ STO with the ratio $c$ of La to Al of 83% (sample ref-A) and 101% (sample ref-B). (c) Temperature dependences of the mobility and the carrier density measured for sample ref-A ($c$ = 83%). (d) Temperature dependence of the sheet resistance of the LSMO (30 u.c.)/ LAO(2 u.c.)/ STO samples with a 2DEG (sample A, $c$ = 83%) and without a 2DEG (sample B, $c$ = 101 %). The dotted curve is the sheet resistance reported for LSMO/STO, which is reproduced from Ref. [25] assuming the film thickness to be the same as that of our LSMO layer.

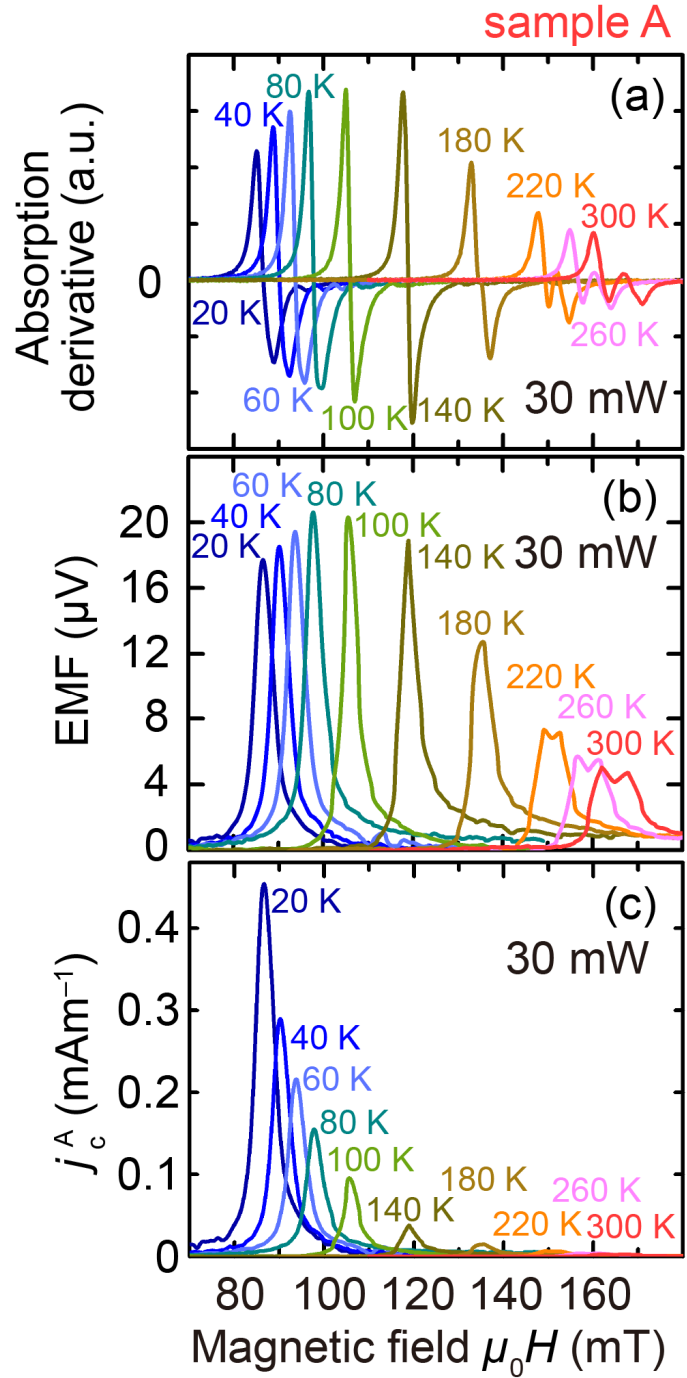

FIG. 3. (a)–(c) Magnetic-field $\mu_0 H$ dependences of the microwave absorption derivative (a), EMF (b), and $j_c^A$ (c) measured for sample A (LSMO/LAO/STO with a 2DEG) at various temperatures. The used microwave power is 30 mW.

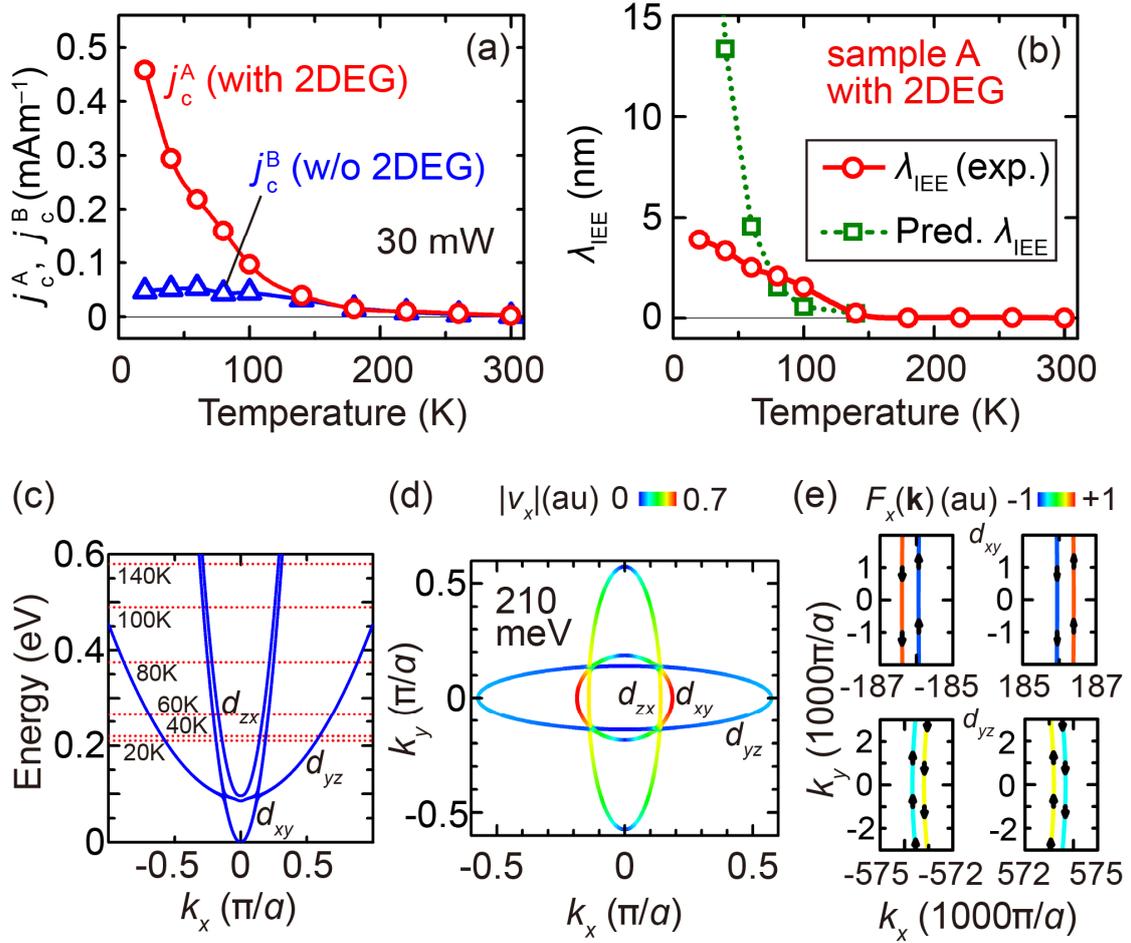

FIG. 4. (a) Comparison between $j_c^A$ (with a 2DEG) and $j_c^B$ (without a 2DEG) as a function of temperature. (b) Temperature dependences of experimental $\lambda_{IEE}$ and predicted $\lambda_{IEE}$. (c) Dispersion relation of the 2DEG at the LAO/STO interface (blue curves). The red dotted lines, from bottom to top, are the estimated $E_F$ positions at 20, 40, 60, 80, 100, and 140 K in our sample. (d) Fermi surface of the 2DEG when the carrier density is $2.1 \times 10^{14}$ cm$^{-2}$ and $E_F = 210$ meV, which corresponds to the measurement condition of the IEE at 20 K. The color scale represents the absolute value of the group velocity in the $x$ direction. (e) Details of the Fermi surface with the spin orientations (see arrows). The color scale represents $F_x(\mathbf{k})$, which is an indicator of the contribution of each state to the electron flow when a spin current is injected.

**Supplemental Material for**

**Efficient intrinsic spin-to-charge current conversion in an all-epitaxial single-crystal perovskite-oxide heterostructure of La$_{0.67}$Sr$_{0.33}$MnO$_3$/LaAlO$_3$/SrTiO$_3$**

Shinobu Ohya,[1,2,3*] Daisei Araki,[1] Le Duc Anh,[1,2] Shingo Kaneta,[1] Munetoshi Seki,[1,3] Hitoshi Tabata,[1,3] and Masaaki Tanaka[1,3]

[1]*Department of Electrical Engineering and Information Systems, The University of Tokyo, 7-3-1 Hongo, Bunkyo-ku, Tokyo 113-8656, Japan*
[2]*Institute of Engineering Innovation, Graduate School of Engineering, The University of Tokyo, 7-3-1 Hongo, Bunkyo-ku, Tokyo 113-8656, Japan*
[3]*Center for Spintronics Research Network (CSRN), Graduate School of Engineering, The University of Tokyo, 7-3-1 Hongo, Bunkyo-ku, Tokyo 113-8656, Japan*

## 1. Influence of the thermal effects

The influence of the Seebeck effect on the electromotive force (EMF) is negligibly small, which can be understood from the linear microwave-power dependence of the electromotive force (EMF) measured for sample A (Fig. S1). The EMF signal induced by the Seebeck effect would be proportional to $S\Delta T$, where $S$ is the Seebeck coefficient of the sample and $\Delta T$ is the temperature difference induced in the sample by microwave irradiation. Because $S$ strongly depends on the temperature [ 29 , 30 ], the power dependence of the EMF would not be linear if it were affected by the Seebeck effect.

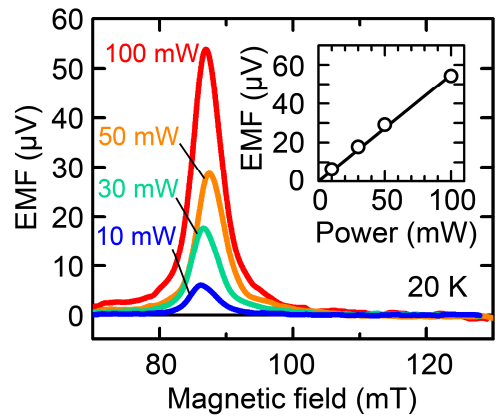

FIG. S1. Magnetic-field $\mu_0 H$ dependence of the EMF measured for sample A with various microwave powers at 20 K. The inset shows the value of the EMF peak as a function of the microwave power at 20 K.

The anomalous Nernst effect and longitudinal spin Seebeck effect generally have been reported to decrease to zero with decreasing temperature [31,32], which is clearly different from our result of the increase in the EMF with a decrease in temperature. Therefore, we can eliminate the influence of the thermal effects on the EMF.

## 2. Derivation of the symmetric component of the magnetic-field dependence of the EMF

To extract the signal of the inverse Edelstein effect (IEE) from the EMF, we need to separate the EMF-$H$ curves ($\mu_0 H$ is a magnetic field) into a symmetric component $V_S F_S(H)$ and an anti-symmetric (diffusive) component $V_A F_A(H)$ [33], where $F_S(H)$ and $F_A(H)$ represent the Lorentzian and anti-Lorentzian functions, respectively. $V_S$ and $V_A$ are those coefficients. It is known that the planar Hall effect (PHE), anomalous Hall effect (AHE), and inverse Edelstein effect (IEE) are incorporated in $V_S F_S(H)$, while $V_A F_A(H)$ is composed only of the PHE and AHE. In Fig. S2, we show the example of a fitting curve of $V_S F_S(H) + V_A F_A(H)$ with the experimental data measured for sample A at 20 K. We see that the EMF-$H$ curve is almost dominated by a symmetric component.

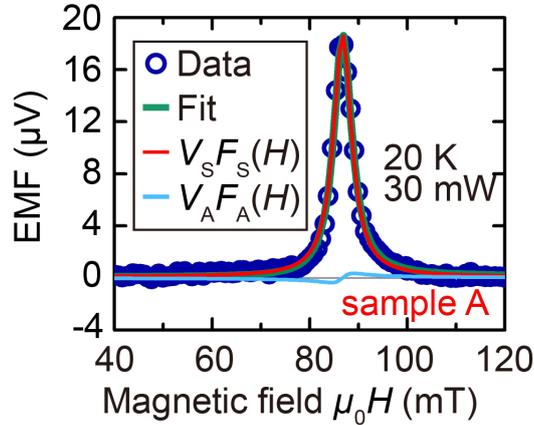

Fig. S2. Example of the fitting (green line) to the EMF signal measured at 20 K for sample A as a function of $\mu_0 H$. The fitting curve is composed of the symmetric [$V_S F_S(H)$] (red) and antisymmetric [$V_A F_A(H)$] (pale blue) curves.

## 3. Spin pumping data of sample B

In Figs. S3(a)–S3(c), we show the results of the ferromagnetic resonance (FMR) and EMF measurements for sample B (LSMO/ LAO/ STO), which does not have a 2DEG and thus only shows AHE and PHE signals originating from LSMO. The generated sheet current density $j_c^B$ in sample B is much smaller than that obtained for sample A ($j_c^A$) [see Fig. 3(c) in the main text], meaning that the obtained EMF in sample A is mainly attributed to the IEE and that AHE and PHE are very small.

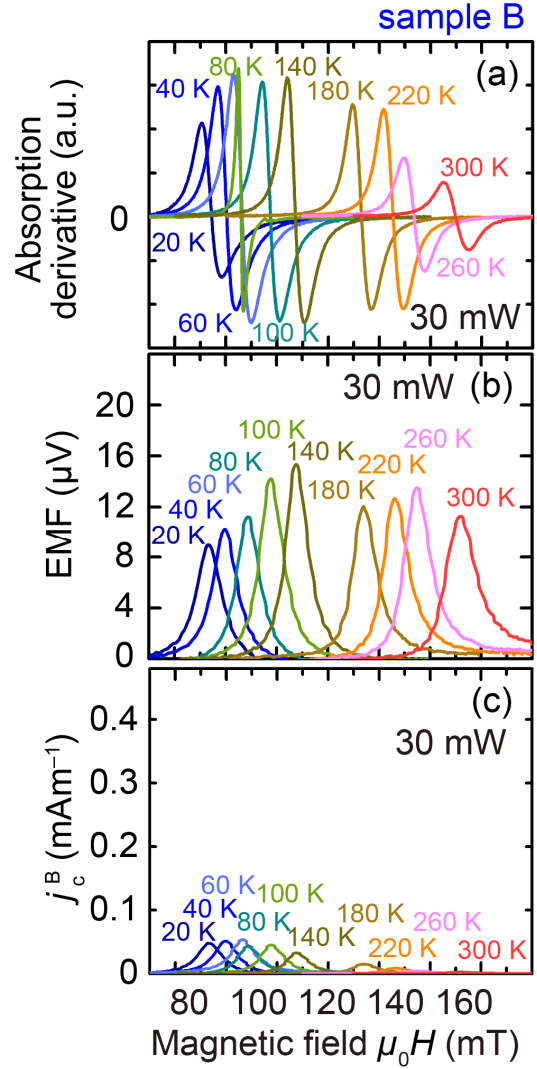

Fig. S3. (a)–(c) Magnetic-field $\mu_0 H$ dependences of the microwave absorption derivative (a), EMF (b), and $j_c^B$ (c) measured for sample B (LSMO/LAO/STO without a 2DEG) at various temperatures. The used microwave power is 30 mW. In (c), the scale of the vertical axis is the same as that in Fig. 3(c) of the main text.

## 4. Estimation of the spin current

We estimated the spin-current density $j_s$ in sample A using

$$j_s = \frac{g^{\uparrow\downarrow}\gamma^2(\mu_0 h_{rf})^2\hbar\left(4\pi M_s\gamma + \sqrt{(4\pi M_s\gamma)^2 + 4\omega^2}\right)}{8\pi\alpha_A^2[(4\pi M_s\gamma)^2 + 4\omega^2]}\left(\frac{2e}{\hbar}\right), \quad (S4.1)$$

where $\hbar$ is the Dirac constant, $\omega$ is the angular frequency of the microwave, $e$ is the elementary charge, $h_{rf}$ is the microwave magnetic field, $\gamma$ is the gyromagnetic ratio in LSMO, $\alpha_A$ is the Gilbert damping constant, and $M_s$ is the saturation magnetization of LSMO [34]. $g^{\uparrow\downarrow}$ is the real part of the spin-mixing conductance given by

$$g^{\uparrow\downarrow} = 4\pi M_s\gamma d_{LSMO}(\alpha_A - \alpha_i), \quad (S4.2)$$

where $d_{LSMO}$ is the thickness of LSMO (12 nm), and $\alpha_i$ is the intrinsic Gilbert damping constant of LSMO with no spin current generation. Due to the spin current generation from LSMO in sample A, $\alpha_A$ is larger than $\alpha_i$. $\alpha_A$ and $\alpha_i$ are obtained by

$$\alpha_\xi = \frac{\sqrt{3}}{2}\frac{g\mu_B}{2\pi f}\frac{1}{\hbar}\Delta H_\xi \quad (\xi = A, i), \quad (S4.3)$$

where $f$ is the frequency of the microwave, $\mu_B$ is the Bohr magneton, and $g$ is the effective electron $g$-factor (1.95 for LSMO [35]). $\Delta H_A$ and $\Delta H_i$ are the experimental FMR spectral linewidths for sample A and intrinsic LSMO (*e.g.* no spin injection), respectively. The obtained temperature dependence of $\alpha_A$ is shown in Fig. S4(a). Here, we set $\alpha_i$ to be $1.57\times10^{-3}$, which was reported for a high-quality LSMO film in Ref. [36]. Depending on the difference of the crystal quality of the LSMO layers, this method may overestimate $j_s$ and thus underestimate $\lambda_{IEE}$ in our study. We have obtained $g^{\uparrow\downarrow} = 73.5$ nm$^{-2}$ and $j_s = 1.06\times10^5$ Am$^{-2}$ at 20 K. The obtained temperature dependence of $j_s$ is shown in Fig. S4(b).

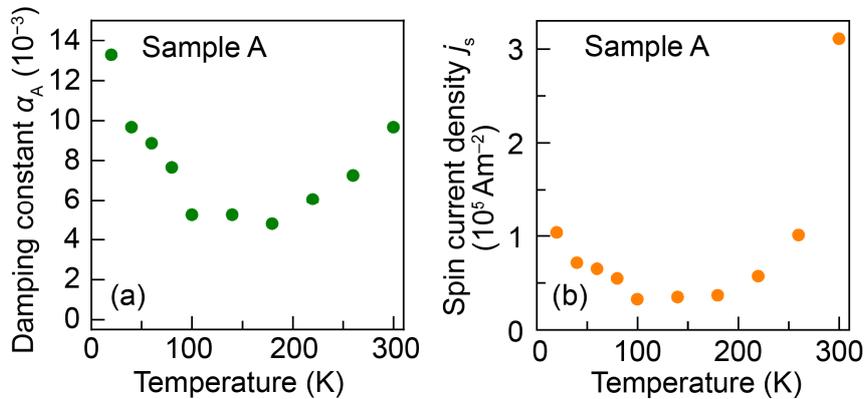

Fig. S4. (a) Obtained damping constant $\alpha_A$ (a) and $j_s$ (b) as a function of temperature.

## 5. Calculation of the band structure and the Fermi surface

Following Ref. [37], we calculated the band structure of the LAO/STO interface assuming that the Hamiltonian is a sum of the nearest-neighbor hopping with on-site interaction, $H_0$, the atomic spin-orbit coupling, $H_{ASO}$, and the interorbital nearest-neighbor hopping, $H_a$. Using the six $t_{2g}$ orbitals of STO, $d_{yz\uparrow}$, $d_{yz\downarrow}$, $d_{zx\uparrow}$, $d_{zx\downarrow}$, $d_{xy\uparrow}$, and $d_{xy\downarrow}$, as basis functions, $H_0$, $H_{ASO}$, and $H_a$ are written as

$$H_0 = \begin{pmatrix} \dfrac{\hbar^2 k_x^2}{2m_h} + \dfrac{\hbar^2 k_y^2}{2m_l} & 0 & 0 \\ 0 & \dfrac{\hbar^2 k_x^2}{2m_l} + \dfrac{\hbar^2 k_y^2}{2m_h} & 0 \\ 0 & 0 & \dfrac{\hbar^2 k_x^2}{2m_l} + \dfrac{\hbar^2 k_y^2}{2m_l} - \Delta_E \end{pmatrix} \otimes \boldsymbol{\sigma}^0, \quad (S5.1)$$

$$H_{ASO} = \Delta_{ASO} \begin{pmatrix} 0 & i\boldsymbol{\sigma}_z & -i\boldsymbol{\sigma}_y \\ -i\boldsymbol{\sigma}_z & 0 & i\boldsymbol{\sigma}_x \\ i\boldsymbol{\sigma}_y & -i\boldsymbol{\sigma}_x & 0 \end{pmatrix}, \quad (S5.2)$$

$$H_a = \Delta_z \begin{pmatrix} 0 & 0 & ik_x \\ 0 & 0 & ik_y \\ -ik_x & -ik_y & 0 \end{pmatrix} \otimes \boldsymbol{\sigma}^0. \quad (S5.3)$$

Here, $\hbar$ is the Dirac constant, $k_x$ and $k_y$ are the $x$- and $y$-direction components of the wave vector $\mathbf{k}$, respectively, and $\boldsymbol{\sigma}^0$ is the identity matrix in spin space. $\boldsymbol{\sigma}_x$, $\boldsymbol{\sigma}_y$, and $\boldsymbol{\sigma}_z$ are the Pauli matrices. $m_l$ and $m_h$ are the effective masses of the light and heavy electrons at the LAO/STO interface, respectively. $\uparrow$ and $\downarrow$ represent spin directions. $\Delta_{ASO}$ and $\Delta_z$ are coefficients that express the magnitudes of $H_{ASO}$ and $H_a$, which were assumed to be 5 and 10 meV, respectively. $\Delta_E$ expresses the energy difference between the $d_{xy}$ and the $d_{zx}$, $d_{yz}$ bands due to the confinement of the wave function in the $z$ direction. We determine $\Delta_E$ to be 90 meV so that the carrier concentration becomes $1.8 \times 10^{13}$ cm$^{-2}$ when the Fermi level $E_F$ position is located at the Lifshitz point (bottom of the $d_{yz}$ band at the $\Gamma$ point), as experimentally confirmed [38]. Here, we calculated the density of states $D(E)$ at the energy $E$ using the Green function method:

$$D(E) = -\frac{1}{\pi} \text{Im} \left[ \lim_{\eta \to 0} \sum_{\mathbf{k}}^{\text{Brillouin Zone}} \frac{1}{E + i\eta - E_{\mathbf{k}}} \right]. \quad (S5.4)$$

where $E_{\mathbf{k}}$ is the dispersion relation of the LAO/STO interface. Using the derived $D(E)$, we obtained the carrier density as a function of $E_F$ (Fig. S5).

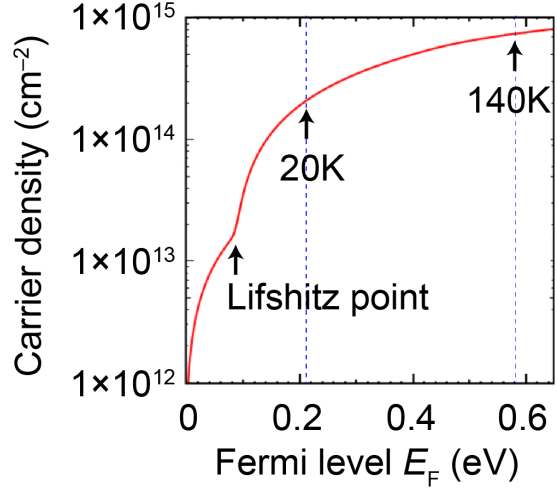

Fig. S5. Calculated carrier density as a function of the Fermi level $E_F$ at the LAO/STO interface. The dotted lines correspond to the carrier concentrations at 20 and 140 K in our sample.

To derive the Fermi surface shown in Figs. 4(d) and 4(e) in the main text, we transformed the Schrödinger equation to Eq. (S5.5), following the method used in Ref. [39].

$$\left[\mathbf{H}^{(2)}k_y^2 + \mathbf{H}^{(1)}(k_x)k_y + \mathbf{H}^{(0)}(k_x) - E\right]\mathbf{C_k} = 0, \tag{S5.5}$$

where $\mathbf{H}^{(2)}$ and $\mathbf{H}^{(1)}(k_x)$ are the coefficients of $k_y^2$ and $k_y$, respectively, $\mathbf{H}^{(0)}$ is the remaining term in the Hamiltonian, $E$ is the electron energy, and $\mathbf{C_k}$ is the eigen vector at $\mathbf{k}$. Equation (S5.5) can be transformed into

$$\begin{pmatrix} \mathbf{0} & \mathbf{1} \\ -\left(\mathbf{H}^{(2)}\right)^{-1}\left[\mathbf{H}^{(0)}(k_x) - E\right] & -\left(\mathbf{H}^{(2)}\right)^{-1}\mathbf{H}^{(1)}(k_x) \end{pmatrix}\begin{pmatrix} \mathbf{C_k} \\ k_y\mathbf{C_k} \end{pmatrix} = 0. \tag{S5.6}$$

By solving this eigen equation for each $k_x$ at $E = E_F$, we obtained $k_y$, $\mathbf{C_k}$, and thus the Fermi surface. The spin direction $\langle\boldsymbol{\sigma}\rangle$, where $\boldsymbol{\sigma} = \left(\boldsymbol{\sigma}_x, \boldsymbol{\sigma}_y, \boldsymbol{\sigma}_z\right)$, at each $\mathbf{k}$ was calculated using $\langle\boldsymbol{\sigma}\rangle = \mathbf{C_k^\dagger}\boldsymbol{\sigma}\mathbf{C_k}$.

## 6. Calculation of the sheet current density and non-equilibrium spin density

Using Eqs. (1)–(3) in the main text, we calculated $j_c^{2D}/\delta s$ as shown in Fig. S6, where $j_c^{2D}$ is the sheet current density induced by IEE and $\delta s$ is the non-equilibrium spin density defined by Eq. (3) in the main text. When $E_F > 0.2$ eV, $j_c^{2D}/\delta s$ is nearly constant as explained in the main text. At around $E_F$=0.1 eV, we see a sharp peak, which corresponds to the Lifshitz point. If we can use this region by controlling the carrier concentration, a large inverse Edelstein length $\lambda_{IEE}$ is expected.

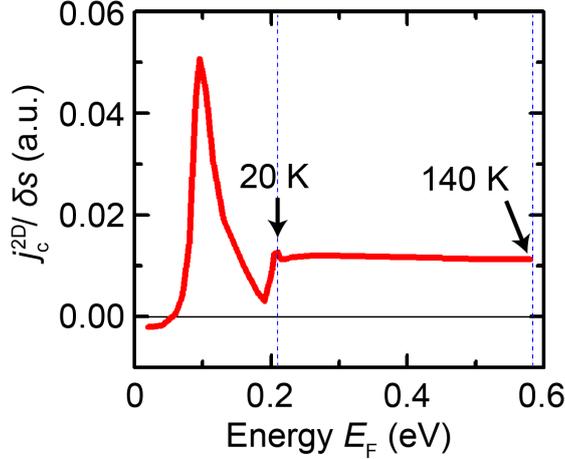

Fig. S6. Calculated $j_c^{2D}/\delta s$ as a function of $E$. The dotted lines correspond to the carrier concentrations at 20 and 140 K in our sample (see Fig. S5).

## 7. Estimation of the relaxation time $\tau$

In the same way as the derivations of Eqs. (1) and (2) in the main text, the following expression between $j_c^{2D}$ and the electric field $F$ in the $x$ direction is obtained.

$$j_c^{2D} = \sum_n \frac{e^2}{4\pi^2\hbar} \int^{SF_n} F\tau(|\mathbf{k}|)v_x(\mathbf{k})\frac{v_x(\mathbf{k})}{|\mathbf{v}(\mathbf{k})|}dS_F, \tag{S6.1}$$

where $e$ is the free electron charge, $dS_F$ is the infinitesimal area (= length in two dimensions) of the Fermi surface, $F$ is the strength of the electric field in the $x$ direction, $v_x(\mathbf{k})$ is the $x$ direction component of the group velocity $\mathbf{v}(\mathbf{k})$, and $n$ is the index of each Fermi surface $SF_n$. Using Eq. (S6.1) and the experimental conductivity of the two-dimensional channel at the LAO/STO interface [see Fig. 2(d), where we neglect the small conductance of LSMO], we can obtain the relaxation time $\tau$, which is expressed by

$$\tau = \left[ \sum_n \int^{\mathrm{FS}_n} \tau(|\mathbf{k}|) \mathrm{d}S_{\mathrm{F}} \right] \Big/ \left( \sum_n \int^{\mathrm{FS}_n} \mathrm{d}S_{\mathrm{F}} \right). \tag{S6.2}$$

The obtained $\tau$ at each temperature is shown in Fig. S7.

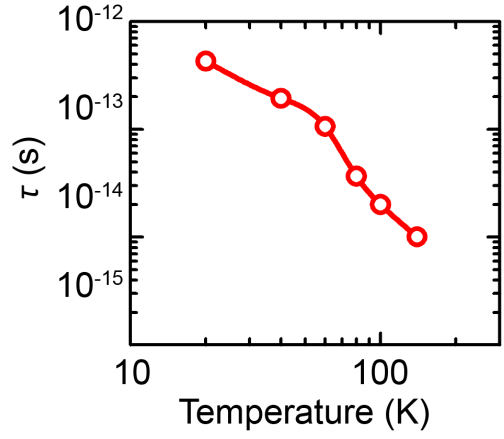

Fig. S7. Obtained temperature dependence of $\tau$.